\documentclass{sn-jnl}
\usepackage[round,authoryear]{natbib}

\let\cite\citep

\usepackage{graphicx}%
\usepackage{multirow}%
\usepackage{amsmath,amssymb,amsfonts}%
\usepackage{amsthm}%
\usepackage{mathrsfs}%
\usepackage[title]{appendix}%
\usepackage{xcolor}%
\usepackage{textcomp}%
\usepackage{manyfoot}%
\usepackage{booktabs}%
\usepackage{algorithm}%
\usepackage{algorithmicx}%
\usepackage{algpseudocode}%
\usepackage{listings}%
\usepackage{siunitx} 
\newcommand{\rev}[1]{\textcolor{black}{#1}}

\usepackage{caption}
\usepackage{subcaption}
\usepackage{svg}
\usepackage{gensymb}
\usepackage{pgfplots}
\pgfplotsset{compat=1.18}
\usepackage{placeins}
\usepackage[colorinlistoftodos]{todonotes}

\theoremstyle{thmstyleone}%
\theoremstyle{thmstyletwo}%

\theoremstyle{thmstylethree}%

\begin{document}

\title[]{Numerical Optimization of \rev{Planar} Nozzle Shapes for Fused Deposition Modeling}


\author*[1]{\fnm{Steffen} \sur{Tillmann}}\email{tillmann@cats.rwth-aachen.de}

\author[1]{\fnm{Felipe A.} \sur{Gonz\'alez}}\email{gonzalez@cats.rwth-aachen.de}

 \author[2,1]{\fnm{Stefanie} \sur{Elgeti}}\email{stefanie.elgeti@tuwien.ac.at}

\affil[1]{\orgdiv{Chair for Computational Analysis of Technical Systems}, \orgname{RWTH Aachen University}, \orgaddress{\street{Schinkelstraße 2}, \city{Aachen}, \postcode{52062}, \country{Germany}}}

\affil[2]{\orgdiv{Institute of Lightweight Design and Structural Biomechanics}, \orgname{TU Wien}, \orgaddress{\street{Gumpendorfer Straße 7}, \city{Vienna}, \postcode{A-1060}, \country{Austria,}}}


\abstract{
\textbf{Purpose}: In fused deposition modeling (FDM), the nozzle plays a critical role in enabling high printing speeds while maintaining precision. Despite its importance, most applications still rely on standard nozzle designs. This work investigates the influence of nozzle geometry on pressure loss inside the nozzle, a key factor in high-speed printing performance. \\
\textbf{Design/methodology/approach}: We focus on optimizing the nozzle shape to minimize the pressure loss and establish a framework that allows both simple angle-based optimization and more advanced spline-based parametrization. To model the polymer melt flow, we use a Giesekus model to account for viscoelastic effects. \\
\textbf{Findings}: For angle-based optimization, the pressure-loss objective exhibits two local minima: one associated with smooth flow and another with pronounced recirculation regions inside the nozzle. While the latter yields a lower pressure drop, such flow patterns are generally undesirable due to increased residence times and the associated risk of material degradation and nozzle clogging. The spline-based parametrization results in only marginal additional reductions in pressure loss compared to angle optimization, while decreasing the manufacturability of the nozzle considerably.\\
\textbf{Originality/value}: This paper presents a comparative study of FDM nozzle shape optimization using a Giesekus model. We introduce a flexible optimization framework that accommodates both simple and advanced geometric parametrizations. \rev{The main contribution is the systematic comparison between angle- and spline-based parametrizations across materials and extrusion velocities, showing that most of the achievable pressure-loss reduction is already captured by the simpler and more manufacture-ready angle optimization.} 
}

%




\maketitle

\section{Introduction}\label{sec1}
With recent technological advances, 3D printing has become a more commonly used manufacturing technique in many fields. While it was initially employed primarily for prototyping, it is now also used for the production of final parts. Among 3D printing technologies, Fused Deposition Modeling (FDM) is particularly widely used due to its low cost, ease of operation, and broad material availability. \cite{paul2021finite, cano2021fused} In the initial stage of FDM, the thermoplastic filament is driven into the hotend by feed rollers, where it is heated above its glass transition and melting temperatures. The polymer transitions into a viscoelastic melt and is forced through a convergent nozzle, where the reduction in cross-section induces an increase in flow velocity and shear rate. Upon deposition onto the build plate, the extrudate undergoes rapid cooling and solidification, governed by heat conduction to the substrate and convective heat transfer to the surrounding environment \cite{hiemenz20113d}. The most commonly used materials are polymers, but other materials such as metals or ceramics are possible \cite{solomon2021review}. In this paper, we will focus only on polymers. \\
To reduce manufacturing costs, printing time must be minimized by increasing the printing speed. However, higher printing speeds result in greater pressure losses within the nozzle, which in turn demand higher feeding forces on the filament \cite{schuller2024optimal, haleem2017mathematical}. Higher printing speeds can furthermore adversely influence the mechanical properties of the produced parts \cite{vzarko2017influence, miazio2019impact, ansari2021effect}. \\
One approach to improve the printing quality is to optimize the process parameters. In \cite{shirmohammadi2021optimization, pereira2023multiobjective}, parameters such as nozzle temperature, layer thickness, printing speed, and nozzle diameter were optimized to improve the mechanical properties of the produced part, such as reducing surface roughness. Further studies have focused on optimizing printing speed  \cite{mushtaq2023investigation} or on performing multi-objective optimization \cite{nguyen2020single}. \\
Most of the aforementioned studies rely on experimental investigations. However, simulations are essential for understanding polymer melt flow in the nozzle, since obtaining high-resolution flow field measurements inside the nozzle is not feasible. Several works have simulated the flow field within the nozzle \cite{nzebuka2022, hofstaetter2015simulation, hajili2024computational}, while others combined simulations with experimental validation \cite{hira2022numerical}. In \cite{serdeczny2020experimental}, experimental measurements of feeding force and pressure loss under different operating conditions were compared against analytical models. \\
The performed studies have mainly focused on determining the characteristics of the polymer melt, such as the flow field and the pressure drop, at the first stage of FDM. These quantities are relevant in the following way: At high extrusion speed, an uneven temperature at the nozzle exit would lead to insufficient melting and a potential risk of clogging. 
Furthermore, the pressure drop is directly correlated to the feeding force, which ultimately determines the maximum manufacturing speed \cite{xu2023}.\\
In literature, the molten polymer is commonly modeled as a generalized Newtonian fluid (GNF) with a time-temperature superposition approach, which accounts for the temperature-dependent viscosity and shear-thinning behavior \cite{serdeczny2020, zhang2023, pigeonneau2020, nzebuka2022, ufodike2022, marion2023, marion2024}.
The GNF model proved to be efficient and accurate in predicting the thermofluid flow inside the hotend at different feeding rates.
However, other researchers prefer viscoelastic models, as these not only describe the viscous flow behavior but also capture the elastic memory effects of polymer melts. Such models are better suited for representing stress relaxation, strain history, and non-linear flow responses, which are important for accurately predicting nozzle flow characteristics \cite{schuller2024assessing, phan2020, serdeczny2022, xu2024}. 
The Weissenberg number (Wi) is used to compare the elastic forces to the viscous forces. Because the polymer melt within the nozzle is dominated by elongational flow, viscoelastic models remain more suitable for accurately capturing the flow behavior \cite{pigeonneau2020, marion2023}.
The material parameters of typical polymers used in FDM, the feeding rate, and the geometry lead to high Wi numbers, which are known to cause numerical difficulties \cite{keunings1986high} and can lead to the formation of large vortices. Continuation techniques are required to achieve convergence at high Wi numbers, where the relaxation time is gradually increased and the mobility factor is decreased. \\
With respect to the pressure loss, it can be observed that most losses occur within the nozzle since the losses through the liquefier are usually insignificant due to its larger diameter \cite{phan2020}.
When GNF and viscoelastic models were compared, the viscoelastic model was shown to be more accurate in predicting the pressure drop in the nozzle and the feeding force \cite{serdeczny2022}.
Viscoelastic flow can also capture elastic instabilities, resulting in increased pressure drops and the formation of recirculation zones along the nozzle walls upstream \cite{schuller2022}.
The presence of such upstream vortices would lead to an inhomogeneous melting distribution, increasing the risk of clogging, as unmelted material could reach the nozzle contraction \cite{atifyardimci1997thermal}. Additionally, the vortices lead to degradation of the polymer that is trapped in these regions \cite{matschinski2021fiber}\rev{, because the prolonged high temperatures and high shear rates cause chain scission of the polymers \cite{pinheiro2004role}.}
Although the polymer's melting is an important aspect of FDM, it is possible to neglect the heating and melting of the polymer in the simulation and assume the polymer is already at working temperature in the nozzle \cite{schuller2024assessing}.\\
As mentioned above, the nozzle plays a crucial part in the process of FDM. The influence of the diameter and temperature of the nozzle was investigated in \cite{kedare20203d}.
In \cite{wang2025optimization}, the contraction angle of the nozzle and the process parameters were optimized together. 
As mentioned before, pressure loss is essential for high printing speeds. In \cite{haleem2017mathematical}, the pressure losses were compared for different contraction angles and outlet diameters. \\
A recent study on the shape optimization of the nozzle geometry using an isothermal viscoelastic fluid has been carried out \cite{schuller2024optimal}. The study demonstrated how the optimal nozzle profile varies for different polymer materials. The nozzle shape was optimized to minimize pressure loss using a spline-based parametrization, which provides greater flexibility in shaping the nozzle.\\
In this paper, we extend this approach by optimizing the nozzle shape for different feeding rates and using different geometry parametrizations. \rev{In contrast to previous work, our emphasis is not on increasing physical model realism, but on comparing how much benefit is obtained from simple angle-based optimization versus a more flexible spline-based parametrization under the same viscoelastic flow model. This allows us to assess the trade-off between pressure-loss reduction, optimizer cost, and manufacturability.} Our procedure first optimizes the contraction angle and subsequently employs a spline-based parametrization to determine the overall optimal nozzle geometry. \\
In general, shape optimization methods can be split into two groups regarding the geometry representation: parametric \cite{azegami2020shape, chapelier2022spline} or nonparametric \cite{hojjat2014vertex, giannakoglou2008adjoint, jameson2003aerodynamic,le2011gradient}. In the case of nonparametric shape optimization, a simulation model capable of the adjoint method is required to compute the shape derivatives of all mesh nodes in a feasible computational time. In cases where the simulation models do not provide objective function derivatives, the only feasible methods are the parametric shape optimization methods, which provide a low-dimensional representation of the geometry \cite{azegami2020shape}. One popular method is free-form deformation \cite{sederberg1986free}, where the geometry is embedded into a box spline, and the control points of the spline serve as the optimization parameters. This works for nozzle-like geometries \cite{tillmann2023shape} as well as more complex geometries \cite{tillmann2024bayesian}. For simpler geometries, it is possible to parametrize the geometry with a spline directly \cite{han2014adaptive, painchaud2006airfoil, schuller2024optimal}. This has the advantage that the spline is better fitted to the geometry and thus allows a more fine-tuned adaptation of the geometry, which is possible for the nozzle geometry. \\
Also, an appropriate optimization algorithm needs to be chosen. In cases where the simulation model does not provide derivatives, derivative-free optimization algorithms are typically used. These algorithms are, for example, explained in \cite{larson2019derivative,conn2009introduction,rios2013derivative}. This paper will use the COBYQA algorithm \cite{rago_thesis, razh_cobyqa}, a trust-region method that can handle all types of constraints. \\
Our paper contributes to a deeper understanding of the flow field and pressure losses within the nozzle for FDM 3D printing applications. We compute the optimal nozzle shape for different operating conditions to minimize pressure loss and provide a general framework for shape optimization that can be applied to any simulation model. The paper is structured as follows: Section \ref{Simulation Model} describes the simulation model and the nozzle geometry. This is followed by a description of our shape optimization framework with different parametrizations in Section \ref{Shape_Optimization}, with the results of the numerical experiments presented in Section \ref{Results}. Finally, we conclude with a summary, discussion, and outlook.

\section{Simulation Model}\label{Simulation Model}

\subsection{Governing Equations Viscoelastic Model}\label{governing_equations}
The governing equations for the viscoelastic model employed in this study are presented in this section. The model is formulated on the basis of the steady, incompressible Navier–Stokes equations. The polymer is assumed to be already fully melted and at working temperature. For a spatial domain $\Omega$, the governing equations of momentum and mass conservation are expressed as:
\begin{equation}
    \rho \mathbf{u} \cdot \nabla \mathbf{u}-\nabla \cdot \boldsymbol{\sigma} =\mathbf{b} \quad\text {in} \quad \Omega,\\
    \label{momentum-equation}
\end{equation}
\begin{equation}
\nabla \cdot \mathbf{u} =0 \quad\text {in} \quad \Omega,\\
\end{equation}
with the density $\rho$, velocity $\mathbf{u}$, external force $\mathbf{b}$, and Cauchy stress tensor $\boldsymbol{\sigma}$. The components of the Cauchy stress tensor depend on the material model. 
To describe viscoelastic material behavior, we employ the incompressible Navier-Stokes equations in conjunction with the Giesekus model \cite{giesekus1982simple}: The model is briefly outlined below, while implementation details, including stabilization, can be found in \cite{wittschieber2022, wittschieberPhd}. We use a logarithmic conformation formulation with the algebraic sub-grid scale method to stabilize the finite element method for viscoelastic flow.
The Cauchy stress tensor in Eq. \ref{momentum-equation} consists of a hydrostatic term $-p \mathbf{I}$, a viscous term $\boldsymbol{\sigma}_{\mathrm{s}}=2 \eta_{\mathrm{s}} \varepsilon(\mathbf{u})$, and additionally the viscoelastic stress $\boldsymbol{\sigma}_{\mathrm{p}}$:
\begin{equation}
\boldsymbol{\sigma}=-p \mathbf{I}+2 \eta_{\mathrm{s}} \varepsilon(\mathbf{u})+\boldsymbol{\sigma}_{\mathrm{p}}.
\end{equation}
Here, $p$ is the pressure, $\eta_{\mathrm{s}}$ is the solvent viscosity, and $\varepsilon$ is the strain rate tensor $\varepsilon(\mathbf{u})=\frac{1}{2}\left(\nabla \mathbf{u}+\nabla \mathbf{u}^{\mathrm{T}}\right)$. The viscoelastic stress is computed with the Giesekus model as the constitutive relation: 
\begin{equation}
\lambda\left(\mathbf{u} \cdot \nabla \boldsymbol{\sigma}_{\mathrm{p}}-\nabla \mathbf{u} \cdot \boldsymbol{\sigma}_{\mathrm{p}}-\boldsymbol{\sigma}_{\mathrm{p}} \cdot(\nabla \mathbf{u})^{\mathrm{T}}\right)+\boldsymbol{\sigma}_{\mathrm{p}} + \frac{\alpha_{G} \lambda}{\eta_{\mathrm{p}}} \boldsymbol{\sigma}_{\mathrm{p}} \cdot \boldsymbol{\sigma}_{\mathrm{p}}=2 \eta_{\mathrm{p}} \varepsilon(\mathbf{u}) \quad \text { in } \quad \Omega,
\end{equation}
with the relaxation time $\lambda$, the mobility factor $\alpha_{G}$, and polymeric viscosity $\eta_{\mathrm{p}}$. In our implementation we specify the total viscosity $\eta$ and the solvent viscosity ratio $\beta$:
\begin{equation}
\eta = \eta_{\mathrm{s}}+\eta_{\mathrm{p}},
\end{equation}

\begin{equation}
\beta = \frac{\eta_{\mathrm{s}}}{\eta_{\mathrm{p}}}.
\end{equation}

\FloatBarrier
\subsection{Model validation}

\begin{figure}[!htbp]
    \centering
    \includegraphics[width=0.5\linewidth]{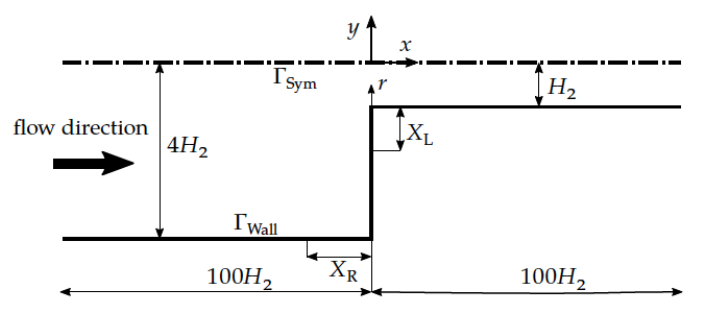}
    \caption{4:1 planar contraction test case geometry. Source: \cite{wittschieberPhd}.}
    \label{fig:contraction}
\end{figure}

The model is validated against reference simulations from the literature, the validation is documented in \cite{wittschieberPhd}; here, only a short summary is given. Two benchmark test cases are considered: flow past a cylinder and a 4:1 planar contraction.
\begin{figure}[!htbp]
    \centering
    \includegraphics[width=0.9\linewidth]{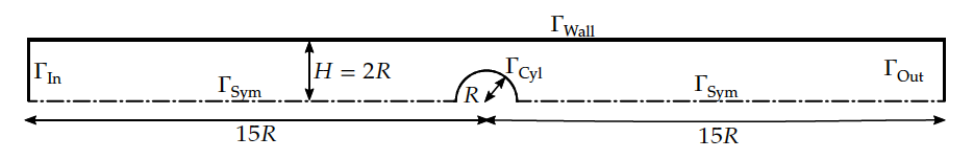}
    \caption{Flow past a cylinder test case geometry. Source: \cite{wittschieberPhd}.}
    \label{fig:cylinder}
\end{figure}
For the flow past a cylinder test case (Figure~\ref{fig:cylinder}), no-slip boundary conditions are imposed on the channel walls and the cylinder surface. Symmetry conditions are applied along the centerline, and the vertical velocity component is set to zero at the outlet. Fully developed velocity and stress profiles are prescribed at both the inlet and the outlet. The material and geometrical parameters are given by
$\beta = 0.59$, $\eta_0 = 1$, $\bar{u} = 2$, $R = 1$, $H = 2$, and $\rho = 0$.
The pressure and drag coefficient are computed for different Weissenberg numbers, defined as $\mathrm{Wi} = \lambda \bar{u} / R \leq 4.0$. The Reynolds number is set to zero by choosing $\rho = 0$.
Simulations are performed for $\alpha_G = 0$, corresponding to the Oldroyd-B model, and for $\alpha_G = 0.1$. The Oldroyd-B results are compared with data from \cite{hulsen2005flow}, \cite{fan1999galerkin}, \cite{afonso2009log}, and \cite{moreno2019logarithmic}. Results obtained with the Giesekus model at $\alpha_G = 0.1$ are compared to those reported in \cite{keith2017ultraweak}. \rev{In Table \ref{tab_validation_cylinder_oldroyd}, the drag coefficient is compared to values from \cite{hulsen2005flow}, showing small relative errors for lower $\mathrm{Wi}$ going up to $1.14\%$ for higher $\mathrm{Wi}$. The results of the comparison from the Giesekus model are shown in Table \ref{tab_validation_cylinder_giesekus}, showing very low relative errors below $0.03\%$ for all tested $\mathrm{Wi}$.}
\begin{center}
\begin{minipage}[t]{0.49\linewidth}
\centering
\footnotesize
\setlength{\tabcolsep}{3pt}
\captionof{table}{\rev{Drag coefficient $C_D$ for the flow-past-a-cylinder benchmark for the Oldroyd-B model ($\alpha_G = 0$) compared to reference \cite{hulsen2005flow}. Source: \cite{wittschieberPhd}}}
\label{tab_validation_cylinder_oldroyd}
\begin{tabular}{@{}c c c c@{}}
\toprule
$\mathrm{Wi}$ & $C_D$ & $C_D^{\mathrm{ref}}$ & Rel.\ error [\%] \\
\midrule
0.5 & 118.883 & 118.781 & 0.086 \\
0.6 & 117.847 & 117.778 & 0.059 \\
0.7 & 117.399 & 117.350 & 0.042 \\
0.8 & 117.436 & 117.380 & 0.048 \\
0.9 & 117.854 & 117.797 & 0.048 \\
1.0 & 118.552 & 118.662 & 0.093 \\
1.1 & 119.481 & 119.740 & 0.216 \\
1.2 & 120.617 & 120.985 & 0.304 \\
1.4 & 123.441 & 124.129 & 0.554 \\
1.5 & 125.098 & 126.022 & 0.733 \\
1.6 & 126.898 & 127.759 & 0.674 \\
1.7 & 128.825 & 130.012 & 0.913 \\
1.8 & 130.874 & 132.024 & 0.871 \\
1.9 & 133.041 & 134.188 & 0.855 \\
2.0 & 135.320 & 136.580 & 0.923 \\
2.2 & 140.183 & 141.801 & 1.141 \\
2.4 & 145.390 & 146.730 & 0.913 \\
2.5 & 148.096 & 149.112 & 0.681 \\
\botrule
\end{tabular}
\end{minipage}
\hfill
\begin{minipage}[t]{0.49\linewidth}
\centering
\footnotesize
\setlength{\tabcolsep}{3pt}
\captionof{table}{\rev{Drag coefficient $C_D$ for the flow-past-a-cylinder benchmark for the Giesekus model ($\alpha_G = 0.1$), compared to reference \cite{keith2017ultraweak}. Source: \cite{wittschieberPhd}}}
\label{tab_validation_cylinder_giesekus}
\begin{tabular}{@{}c c c c@{}}
\toprule
$\mathrm{Wi}$ & $C_D$ & $C_D^{\mathrm{ref}}$ & Rel.\ error [\%] \\
\midrule
0.1 & 125.5528 & 125.5871 & 0.027 \\
0.2 & 117.0816 & 117.1127 & 0.027 \\
0.3 & 111.0703 & 111.0985 & 0.025 \\
0.4 & 106.8292 & 106.8551 & 0.024 \\
0.5 & 103.7089 & 103.7331 & 0.023 \\
0.6 & 101.3187 & 101.3416 & 0.023 \\
0.7 & 99.4263 & 99.4481 & 0.022 \\
0.8 & 97.8882 & 97.9093 & 0.022 \\
0.9 & 96.6113 & 96.6317 & 0.021 \\
1.0 & 95.5326 & 95.5525 & 0.021 \\
\botrule
\end{tabular}
\end{minipage}
\end{center}
The 4:1 planar contraction geometry (Figure~\ref{fig:contraction}) closely resembles the nozzle geometry considered in this work. Parabolic velocity profiles are prescribed at the inlet and outlet, no-slip boundary conditions are imposed on the walls, the pressure is set to zero at the outlet, and symmetry conditions are applied at the top boundary. The parameters used in this test case are
$\beta = 1/9$, $\eta_0 = 1$, $H_2 = 1$, $\overline{u_2} = 1$, and $\rho = 0$.
Simulations are performed for Weissenberg numbers up to $\mathrm{Wi}=\lambda \overline{u_2}/H_2 \leq 12.0$ with $\alpha_G = 0$.
The size of the corner vortex is compared with results from \cite{alves2003benchmark}, \cite{pimenta2017stabilization}, and \cite{niethammer2018numerical}. \rev{In Table \ref{tab_validation_contraction} the results compared to \cite{pimenta2017stabilization} are shown, showing relative errors around $1\%$.}
\rev{Full details of the quantitative error analysis} of the validation study are provided in \cite{wittschieberPhd}.
\begin{table}[!htbp]
\centering
\caption{\rev{Corner vortex size for the 4:1 planar contraction benchmark compared to the reference data from \cite{pimenta2017stabilization}. Source: \cite{wittschieberPhd}}}
\label{tab_validation_contraction}
\begin{tabular}{c c c c}
\toprule
$\mathrm{Wi}$ & $L_c$ & $L_c^{\mathrm{ref}}$ & Rel.\ error [\%] \\
\midrule
0.0 & 1.492 & 1.495 & 0.201 \\
1.0 & 1.364 & 1.365 & 0.073 \\
2.0 & 1.169 & 1.172 & 0.256 \\
3.0 & 0.968 & 0.972 & 0.412 \\
4.0 & 0.787 & 0.788 & 0.127 \\
5.0 & 0.641 & 0.638 & 0.470 \\
6.0 & 0.533 & 0.527 & 1.139 \\
7.0 & 0.458 & 0.453 & 1.104 \\
8.0 & 0.401 & 0.398 & 0.754 \\
9.0 & 0.354 & 0.353 & 0.283 \\
\botrule
\end{tabular}
\end{table}
\rev{Together with the mesh-sensitivity results reported in Section \ref{Mesh study}, this supports
that the numerical uncertainty of the present solver setup is smaller than the improvement by the shape optimization.}
\FloatBarrier
\subsection{Geometry, Boundary Conditions and Material Parameters}
\label{Boundary Conditions}

\begin{figure}[!htbp]
    \centering
    \includegraphics[width=0.9\linewidth,]{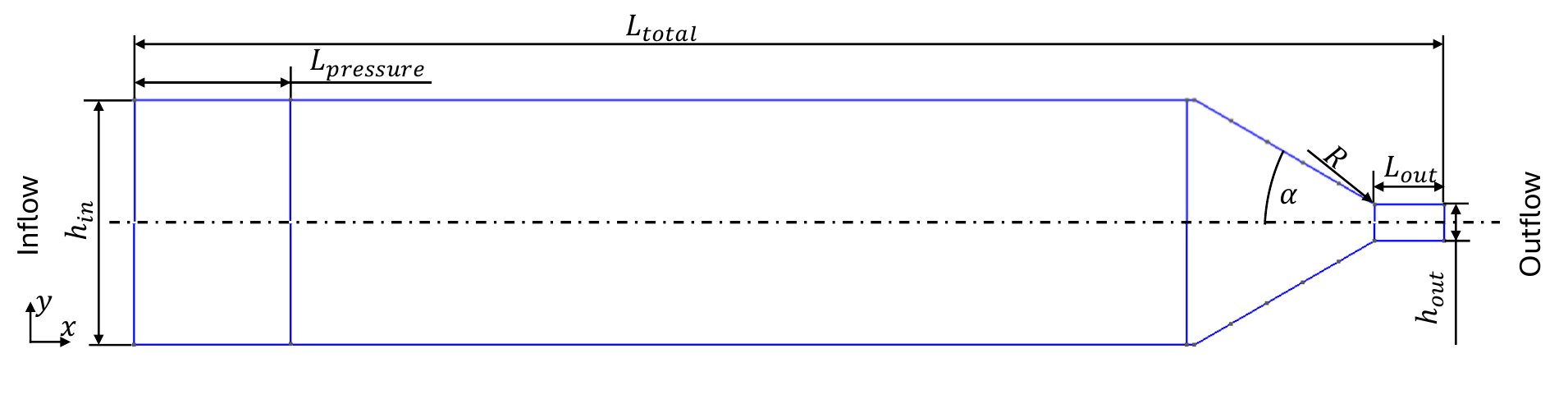}
    \caption{Nozzle geometry. Source: Authors own work}
    \label{fig:Nozzle}
\end{figure}

Figure~\ref{fig:Nozzle} illustrates the nozzle geometry considered in this study. Owing to geometric symmetry, only the upper half of the nozzle is simulated. For reasons of computational efficiency, the present study is limited to a planar two-dimensional geometry. This choice is motivated by the optimization setting, which necessitates a large number of forward-model evaluations. We acknowledge the geometric simplification introduced by this choice; however, when compared with the literature, the results remain qualitatively consistent and are therefore sufficient for the purposes of optimization.
\rev{The objective of the present work is therefore a comparative assessment of geometry
parametrizations for pressure-loss reduction, not a fully realistic representation of
the complete hotend. Axisymmetric and non-isothermal effects are outside the present
scope and are treated as future work.}
At the re-entry corner, a small radius is added for two reasons: (1) It makes the geometry more realistic, as in practice, the nozzle wears out and does not have sharp corners and (2) this radius also helps with numerical stability of the viscoelastic model. The specific values for each length are given in Table~\ref{tab_geometry_experiment}. \\
The following boundary conditions are prescribed: In all cases, we set a parabolic inlet velocity profile, and at the symmetry axis, the velocity in the y-direction is set to zero. The inlet velocity is prescribed such that a desired average extrusion velocity is achieved. At the wall, a no-slip condition is applied. The fluid stresses are prescribed as zero at the inlet, and a symmetry condition is applied at the symmetry axis. At the outlet, the pressure is set to zero.\\
In this study, three polymers commonly used in FDM are considered: Polylactic Acid (PLA), Polyethylene Terephthalate Glycol-modified (PET-G), and a Polyamide co-polymer blend (PA6/66). The material parameters are given in Table~\ref{tab_material}. \rev{The same
Giesekus constitutive model is used for all three materials because literature parameter
sets are available for comparable FDM operating conditions, which enables a
controlled comparison of geometric effects across materials within one consistent viscoelastic
framework.} Note that all polymers are assumed to be fully melted at their respective working temperature (see \cite{schuller2024assessing}).

\begin{table}[!htbp]
\caption{Geometry parameters used for the numerical experiments} 
\label{tab_geometry_experiment}%
\centering
\begin{tabular}{@{}l S l@{}}
\toprule
Parameter & {Value} & {Unit} \\
\midrule
$L_{total}$     & 18     & \si{\milli\meter} \\
$L_{out}$       & 0.9    & \si{\milli\meter} \\
$L_{pressure}$  & 4      & \si{\milli\meter} \\
\rev{$h_{in}$}      & 3.2    & \si{\milli\meter} \\
\rev{$h_{out}$}       & 0.5      & \si{\milli\meter} \\
\rev{$A_{in}/A_{out}$} & 6.4 & \si{-}\\
$R$             & 0.01      & \si{\milli\meter} \\
\bottomrule
\end{tabular}
\end{table}

\begin{table}[!htbp]
\centering
\caption{Material parameters. Data taken from \cite{schuller2024assessing}.}
\label{tab_material}

\begin{tabular}{l c c c}
\toprule
Parameter & {PLA} & {PET-G} & {PA6/66} \\
\midrule
Total viscosity $\eta$ [\si{\pascal\second}]  & 2220 & 600  &  1300 \\
Viscosity ratio $\beta$ [\si{-}]  & 0.15 & 0.15  &  0.15 \\
Relaxation time $\lambda$  [\si{\second}] & 0.209 & 0.355  &  0.518  \\
Mobility factor $\alpha_G$  [\si{-}]   & 0.04 & 0.05 & 0.24  \\
Density $\rho$ [\si{\kilogram\per\meter\cubed}]    & 1250 & 1250 & 1150  \\

\botrule
\end{tabular}
\end{table}
\FloatBarrier
\subsection{Mesh study}
\label{Mesh study}
\begin{figure}[!htbp]
    \centering
    \begin{subfigure}[b]{0.45\textwidth}
        \includegraphics[width=\linewidth]{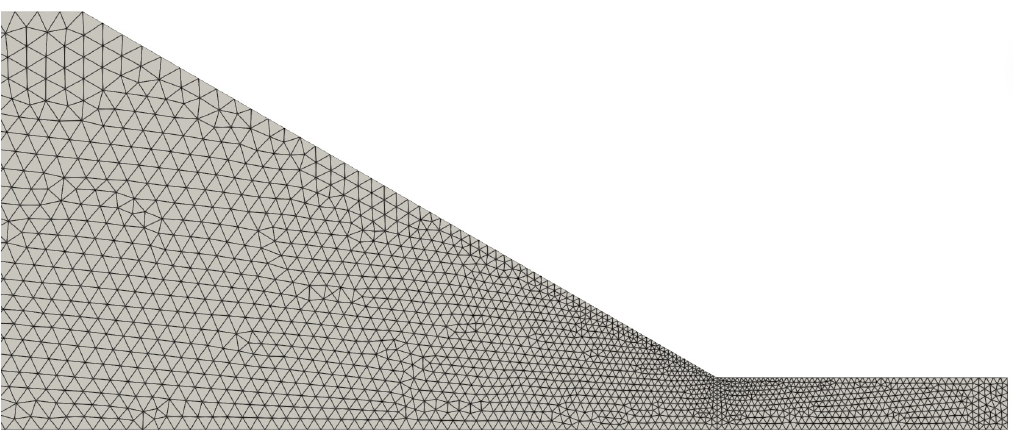}
        \caption{Coarse mesh}
        \label{Coarse mesh study}
    \end{subfigure}
    \hfill
    \begin{subfigure}[b]{0.45\textwidth}
        \includegraphics[width=\linewidth]{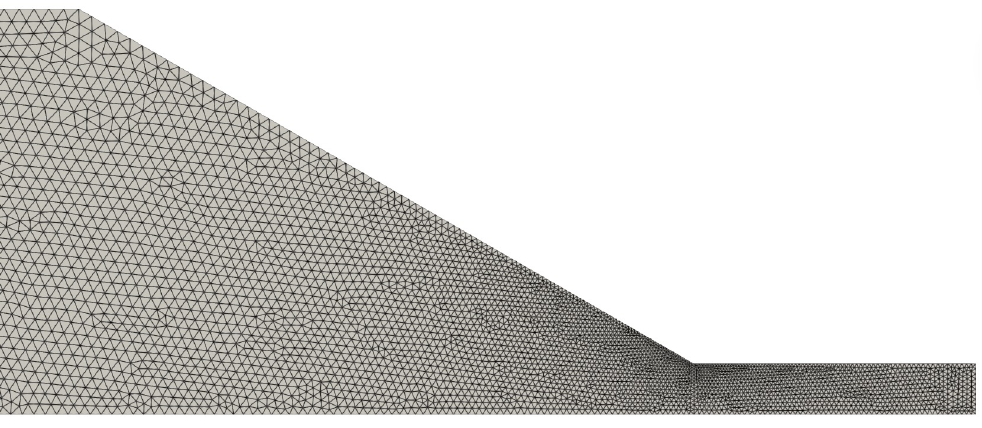}
        \caption{Medium mesh}
        \label{Medium mesh study}
    \end{subfigure}
    \hfill
    \begin{subfigure}[b]{0.45\textwidth}
        \includegraphics[width=\linewidth]{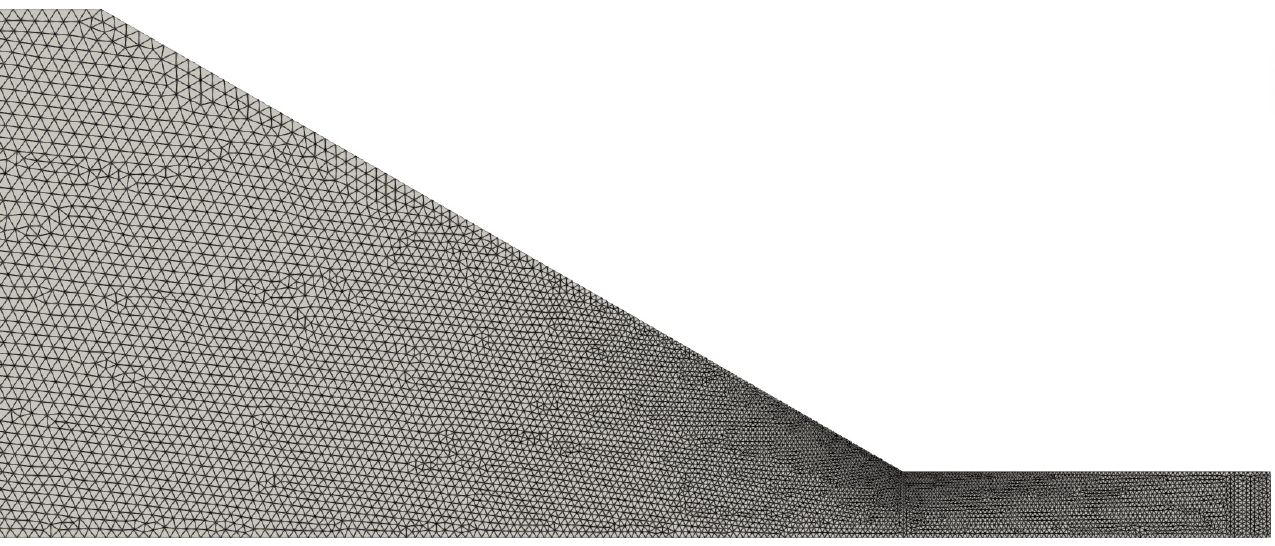}
        \caption{Fine mesh}
        \label{Fine mesh study}
    \end{subfigure} 
    \hfill
    \begin{subfigure}[b]{0.45\textwidth}
        \includegraphics[width=\linewidth]{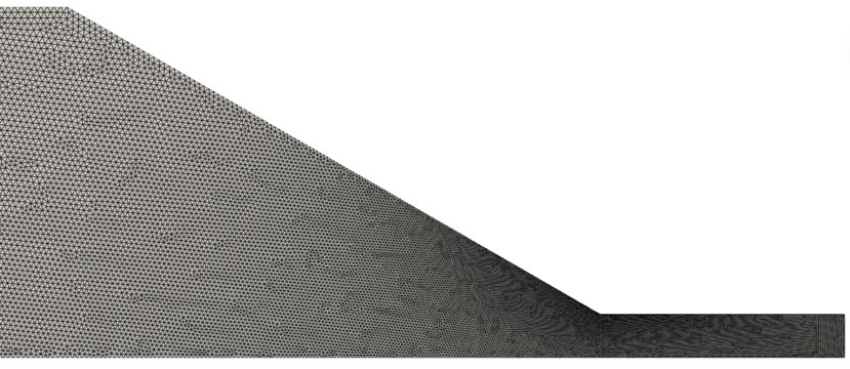}
        \caption{Very fine mesh}
        \label{Very fine mesh study}
    \end{subfigure}

   \caption{Used meshes in the mesh study. Source: Authors own work} 
    \label{mesh study all meshes}
\end{figure}

To obtain mesh-independent results, a mesh refinement study was performed for the selected geometry (Figure~\ref{fig:Nozzle}) with a contraction angle of $\alpha=30\degree$. The mesh is locally refined at the outlet and re-entry corner, as in these regions the largest gradients of the solution variables occur. In the refinement process, the relative refinement pattern is kept unchanged, while the overall mesh size is uniformly reduced (Figure~\ref{mesh study all meshes}). The material used for the mesh study is PET-G and boundary conditions as described in Section~\ref{Boundary Conditions}. The average extrusion velocity is $110$ mm/s. The pressure drop across the nozzle is our quantity of interest. The results, summarized in Table~\ref{tab_mesh_study_results}, show a clear convergence trend as the mesh is refined. While noticeable differences are observed between the coarse (Figure~\ref{Coarse mesh study}) and medium meshes (Figure~\ref{Medium mesh study}), the change in pressure drop between the fine (Figure~\ref{Fine mesh study}) and very fine meshes (Figure~\ref{Very fine mesh study}) is marginal, indicating that the solution is sufficiently resolved at the fine mesh level. \rev{For the selected quantity of interest, the difference between the fine and very fine meshes is approximately 0.07\%, which is substantially smaller than the 4--5\% pressure-loss reductions discussed in Section~\ref{Results}.} In particular, further mesh refinement leads to only minor improvements in accuracy at a substantially increased computational cost. The wall times reported for each mesh correspond to simulations performed on 30 CPU cores. Given that the targeted application involves shape optimization, which requires a large number of repeated flow simulations (up to 100 runs), the fine mesh provides an appropriate compromise between accuracy and efficiency. Therefore, this mesh is deemed sufficient and used for all subsequent numerical experiments.

\begin{table}[!htbp]
\centering
\caption{Mesh study results. Source: Authors own work}
\label{tab_mesh_study_results}

\begin{tabular}{l S S S S}
\toprule
Mesh & {Relative mesh size} & {Number of nodes} & {Pressure drop [MPa]}& {Wall time [hh:mm:ss]} \\
\midrule
Coarse   & 0.30 & 3048  & 0.525969 & \text{00:02:59}\\
Medium   & 0.20 & 6505  & 0.527158 & \text{00:08:24}\\
Fine   & 0.15 & 11390  &  0.528178 & \text{00:17:37}\\
Very Fine     & 0.10 & 24982 & 0.528524 & \text{01:01:32}\\
\botrule
\end{tabular}
\end{table}

\FloatBarrier
\section{Shape Optimization Approach}\label{Shape_Optimization}

\begin{figure}[!htbp]
\begin{subfigure}{0.5\textwidth}
        \includegraphics[width=\linewidth, angle=0]{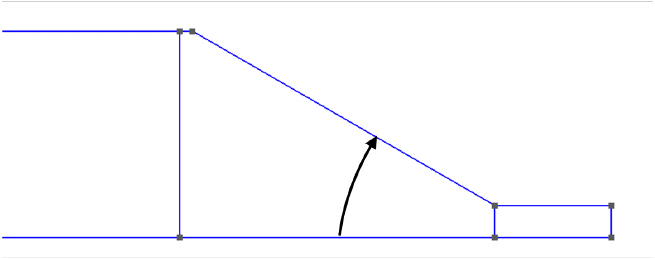}
        \caption{Angle parametrization.}
        \label{fig:Parametrization_a}
\end{subfigure}
\begin{subfigure}{0.5\textwidth}
    \includegraphics[width=\linewidth]{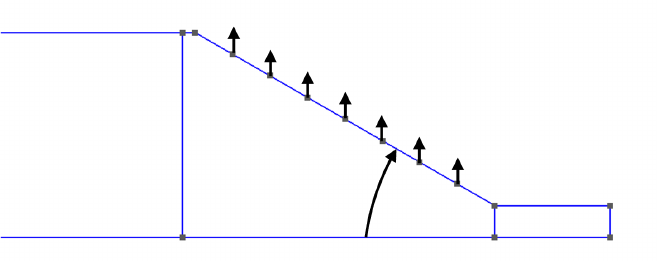}
    \caption{Combined spline parametrization.}
    \label{fig:Parametrization_b}
\end{subfigure}
\caption{The two different geometry parameterizations used. Source: Authors own work}
\label{fig:Parametrization}
\end{figure}

The objective of the shape optimization is to reduce the feeding force by minimizing the pressure loss. To this end, the average pressure is computed at both the inlet and the outlet, and the difference is taken as the pressure loss. Specifically, the pressure at each section is obtained by averaging over its cross-sectional area. At the inlet, the pressure is evaluated slightly downstream, at a distance $L_{pressure}$ from the actual boundary, in order to obtain more consistent values and to avoid disturbances induced by the inflow condition. Our objective function reads: 
\begin{equation}
    J = p_{inlet} - p_{outlet} = \Delta p \:.
    \label{objective function}
\end{equation}
Each average pressure value is computed by integrating the pressure over the corresponding cross-sectional area and normalizing by the area size. To facilitate comparison across different operating conditions, we additionally define the relative improvement in pressure loss and the relative pressure loss, computed with respect to the initial geometry featuring a $30 \degree$ contraction angle.
\begin{equation}
    \text{relative improvement} = 1 - \frac{\Delta p (\text{optimized shape})}{\Delta p (30 \degree \text{ angle})}  \:,
    \label{relative improvement}
\end{equation}
\begin{equation}
    \text{relative pressure loss} = \frac{\Delta p (\text{optimized shape})}{\Delta p (30 \degree \text{ angle})}  \:.
    \label{relative pressure loss}
\end{equation}
We employ parametric shape optimization for two main reasons: it ensures computational efficiency by keeping the number of parameters low, and it incorporates manufacturing constraints. To minimize production costs, the geometry should remain as simple as possible; therefore, in the first step, we restrict the optimization to the contraction angle $\alpha$ (Figure~\ref{fig:Parametrization_a}). \\
To investigate whether increased shape flexibility can lead to improved nozzle designs, we subsequently introduce a spline-based parametrization. A spline parametrization allows for more complex shapes at the cost of higher complexity in the nozzle's manufacturing process. The spline parametrization furthermore introduces new challenges for shape optimization. Not only is there a strong increase in computational complexity due to the much larger number of parameters, but if all control points can move freely in the x- and y-directions, the control points can easily overlap and entangle the mesh. There are different techniques to prevent mesh entangling, such as using different types of constraints. In our case, we restrict the individual movement of the control points in the x-direction. All control points' x-coordinates are controlled by a single parameter that controls the contraction angle, and only the y-coordinates can vary individually (Figure~\ref{fig:Parametrization_b}). In this way, we simultaneously reduced the degrees of freedom to save computational time while ensuring that the mesh remained well-behaved and free of entanglement. Preliminary tests indicated that this approach significantly lowered computational cost while still providing sufficient flexibility in the design, compared to allowing all control points to move in both coordinate directions.
\\
One challenge remains: We need a monotonic decrease in the contraction to maintain reasonable manufacturability. For this, we added an inequality constraint: the y-coordinate of each subsequent control point in the x-direction has to be lower than the previous one.\\
The optimization loop then works as follows. First, the geometry for our simulation model is created according to the initial values of the parametrization. We use the meshing software Gmsh \cite{geuzaine2009gmsh} to create the parametrized geometry and mesh it. We then run the flow simulation with our in-house finite element solver, XNS. Then the pressure values are computed for the objective function, and as the optimization algorithm, we use the COBYQA algorithm \cite{rago_thesis,razh_cobyqa}. \rev{COBYQA is a deterministic, derivative-free
trust-region method that} uses a quadratic approximation of the objective function in every iteration. It is designed explicitly for derivative-free problems and can handle equality and inequality constraints, so it is well-suited for our problem. As we need an inequality constraint for the y-coordinates of the control points to ensure easy manufacturability. The optimization algorithm then computes the values of the geometry parametrization for the next iteration. The primary hyperparameter of the algorithm is the initial search radius, which defines the maximum parameter perturbation from the initial design in the first optimization iteration. 
\rev{Using the fine mesh from Table~\ref{tab_mesh_study_results}, one forward simulation takes about 18 minutes on 30 CPU cores. This corresponds to roughly 4.4 hours for a 15-iteration angle optimization and about 29-47 hours for a 100-160 iteration spline optimization.}
This optimization loop is then run for different materials and extrusion velocities; the results are shown in the next section.\\
\FloatBarrier
\section{Numerical Experiments}\label{Results}
This section contains the numerical experiments: First, the results from the angle optimization are presented, followed by those for the spline-based parametrization. The geometry, boundary conditions, and material parameters are specified in Section~\ref{Boundary Conditions}. Three different materials are used in the numerical experiments: PA6/66, PET-G, and PLA. For the angle optimization, four different extrusion velocities are tested. For the computationally expensive spline-based shape optimization, only an extrusion velocity of 110 mm/s is used.

\FloatBarrier
\subsection{Angle Parametrization}\label{Angle Parametrization}
\begin{figure}[!htbp]
    \centering
    {\centering
      \includegraphics[width=0.5\textwidth]{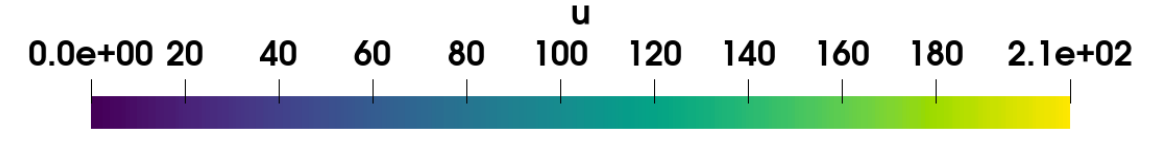} 
      \\
      }
    \begin{subfigure}[b]{0.49\textwidth}
        \includegraphics[width=\linewidth]{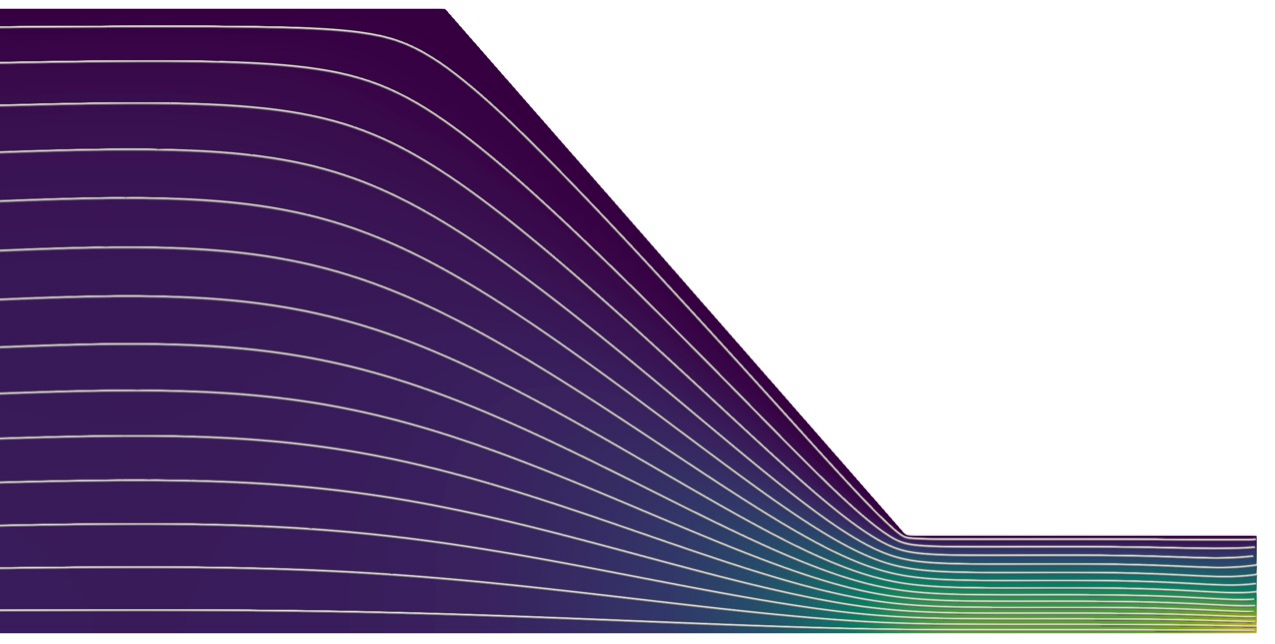}
        \caption{$\alpha = 51.7 \degree,\Delta p = 1.9218 $ MPa}
        \label{fig:PLA-bild1-flow-field}
    \end{subfigure}
    \hfill
    \begin{subfigure}[b]{0.49\textwidth}
        \includegraphics[width=\linewidth]{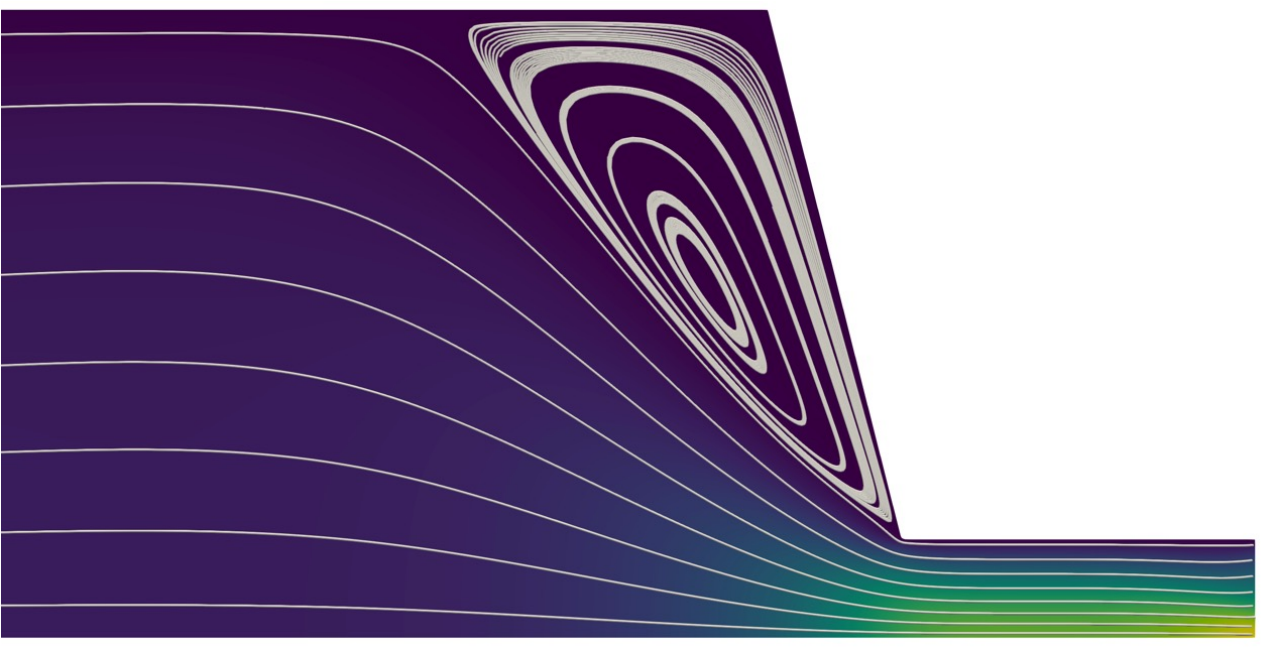}
        \caption{$\alpha = 75.8 \degree,\Delta p = 1.9058$ MPa}
        \label{fig:PLA-bild2-flow-field}
    \end{subfigure}
   \caption{Flow field of the velocity in x-direction (u \rev{[$\si{\milli\meter\per\second}$]}) with the streamlines for the two optimal angles for PLA material. The extrusion velocity is $110$~mm/s. Source: Authors own work} 
   \label{fig:PLA-flow-field-angle}
\end{figure}

\begin{figure}[!htbp]
    \centering
    {\centering
      \includegraphics[width=0.5\textwidth]{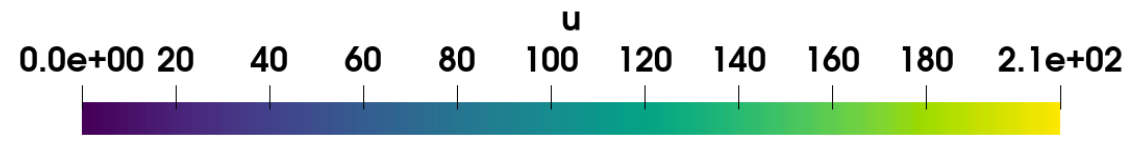} 
      \\
      }
    \begin{subfigure}[b]{0.49\textwidth}
        \includegraphics[width=\linewidth]{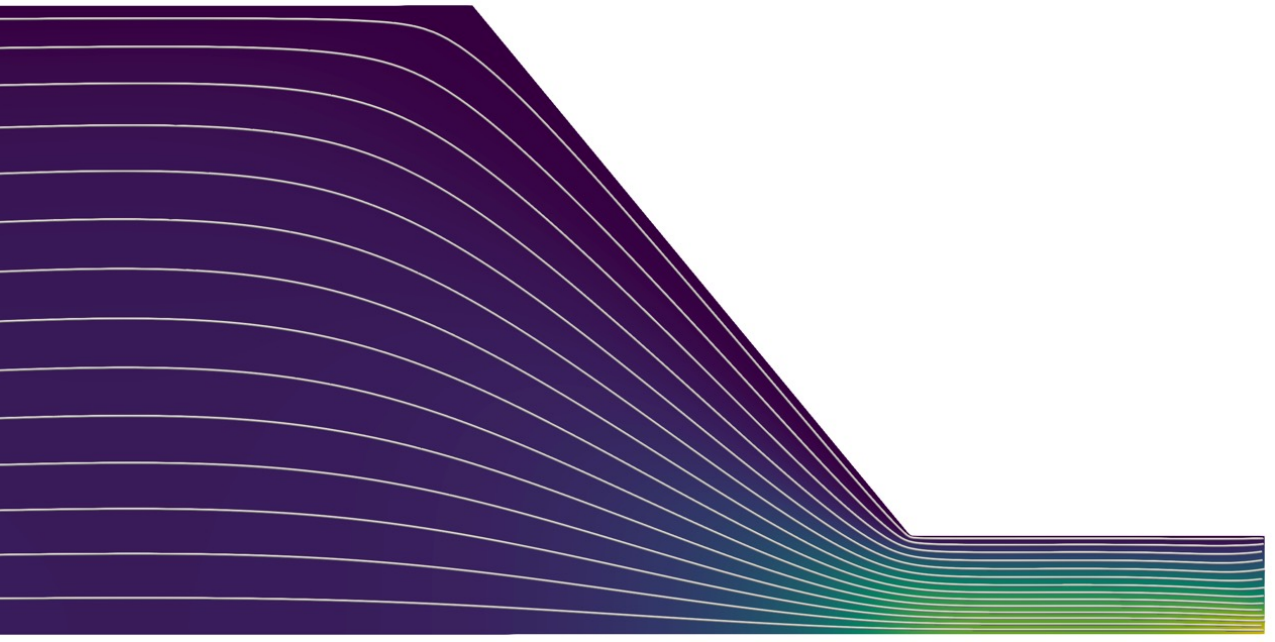}
        \caption{$\alpha = 50.5\degree,\Delta p =  0.5048$ MPa}
        \label{fig:PET-G-bild1-flow-field}
    \end{subfigure}
    \hfill
    \begin{subfigure}[b]{0.49\textwidth}
        \includegraphics[width=\linewidth]{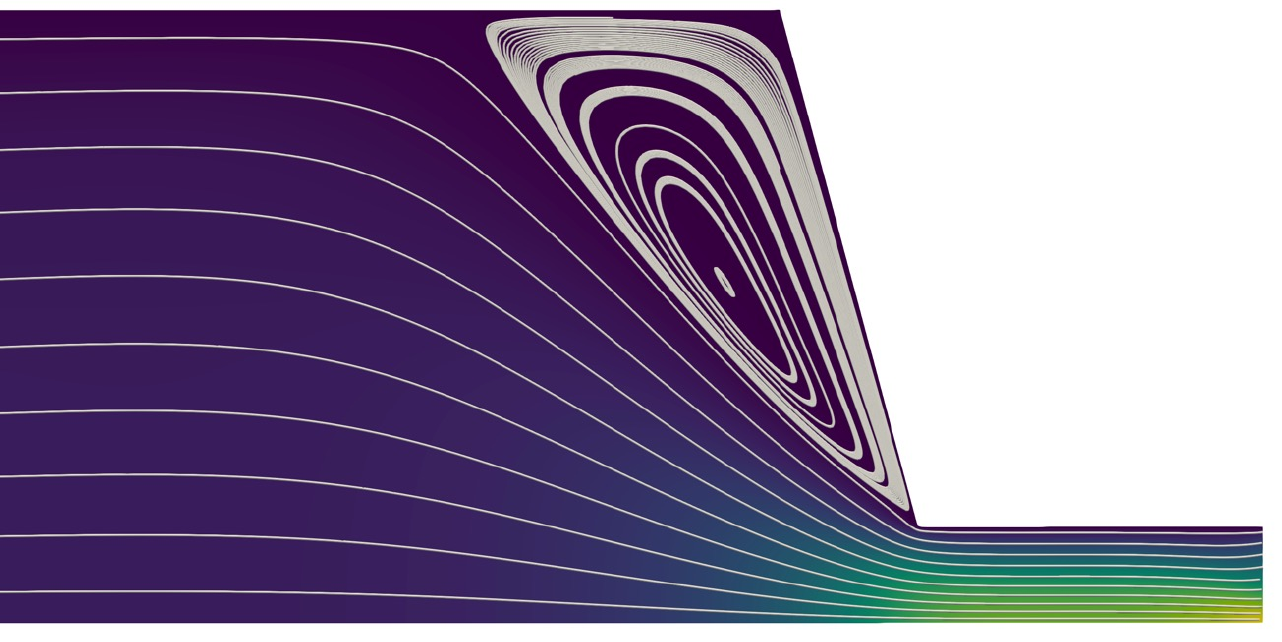}
        \caption{$\alpha = 75.1 \degree,\Delta p = 0.5011$ MPa}
        \label{fig:PET-G-bild2-flow-field}
    \end{subfigure}
   \caption{Flow field of the velocity in x-direction (u \rev{[$\si{\milli\meter\per\second}$]}) with the streamlines for the two optimal angles for PET-G material. The extrusion velocity is $110$~mm/s. Source: Authors own work} 
   \label{fig:PET-G-flow-field-angle}
\end{figure}

\begin{figure}[!htbp]
    \centering
    {\centering
      \includegraphics[width=0.5\textwidth]{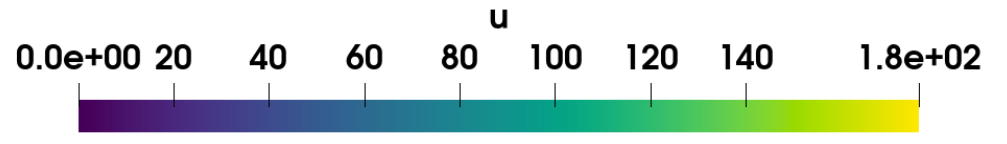} 
      \\
      }
    \begin{subfigure}[b]{0.49\textwidth}
        \includegraphics[width=\linewidth]{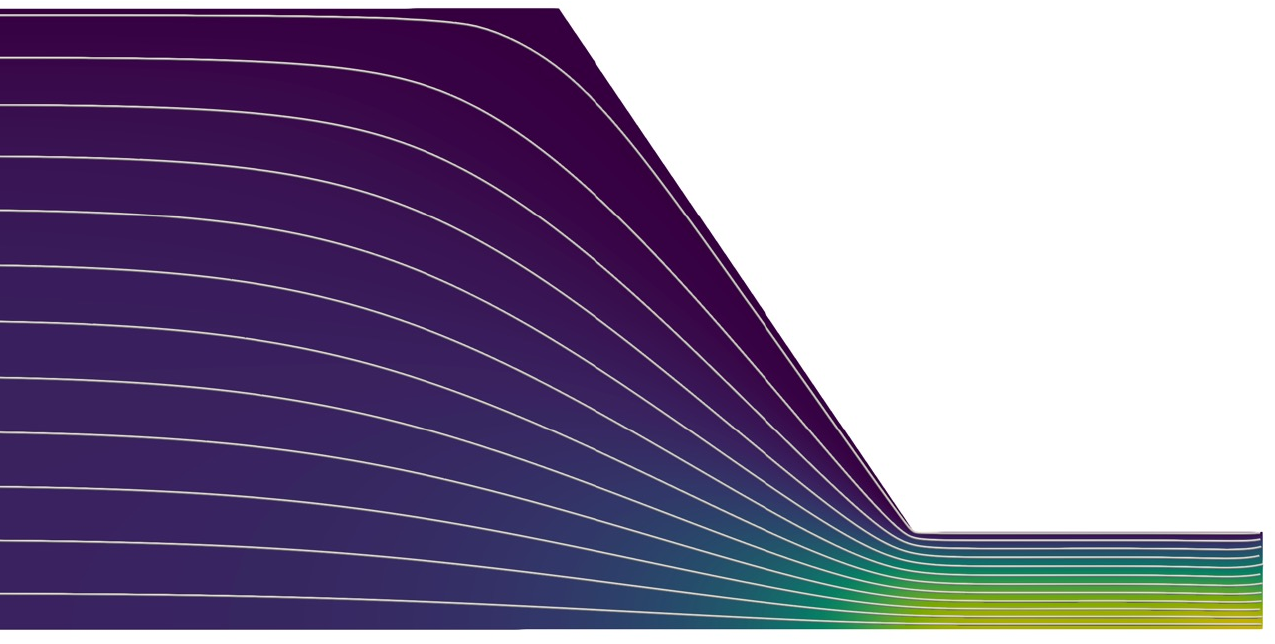}
        \caption{$\alpha = 56.3 \degree,\Delta p = 1.1237 $ MPa}
        \label{fig:PA6-66-bild1-flow-field}
    \end{subfigure}
    \hfill
    \begin{subfigure}[b]{0.49\textwidth}
        \includegraphics[width=\linewidth]{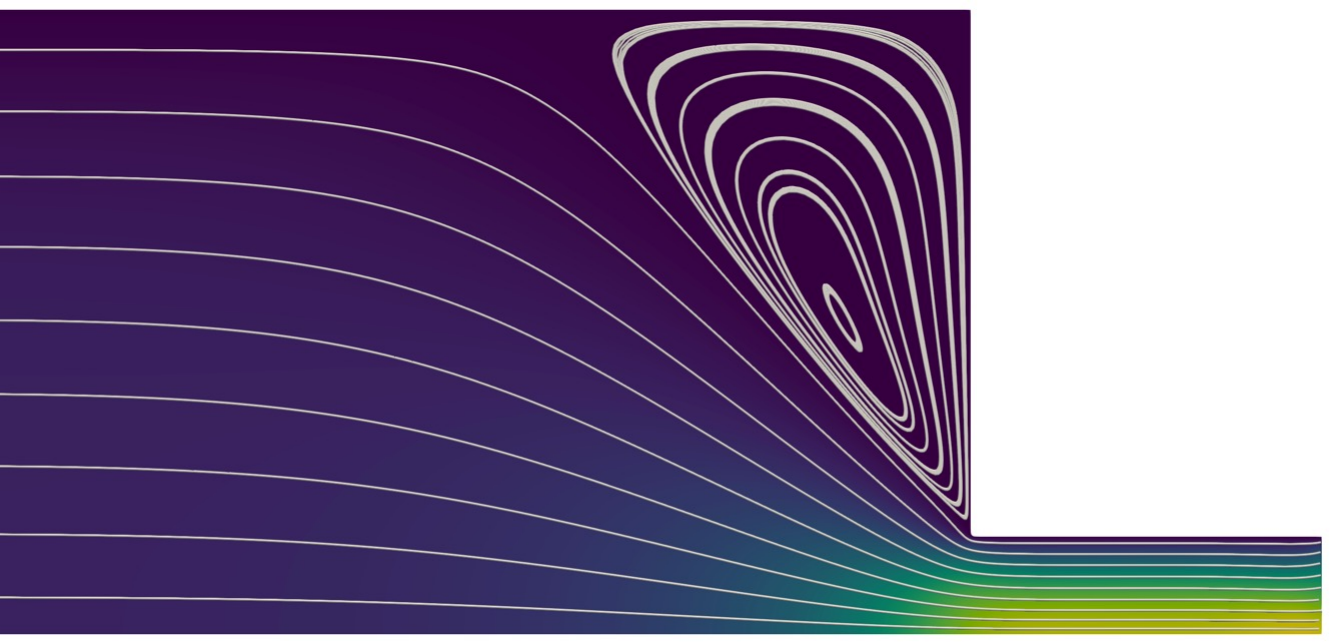}
        \caption{$\alpha = 90 \degree,\Delta p = 1.1153 $ MPa}
        \label{fig:PA6-66-bild2-flow-field}
    \end{subfigure}
   \caption{Flow field of the velocity in x-direction (u \rev{[$\si{\milli\meter\per\second}$]}) with the streamlines for the two optimal angles for PA6/66 material. The extrusion velocity is $110$~mm/s. Source: Authors own work} 
   \label{fig:PA6-66-flow-field-angle}
\end{figure}

We first performed the angle optimization for the extrusion velocity of 110 mm/s for all three materials. For each material, the optimization algorithm found the optimal angle. What we found is that, depending on the starting value, the optimizer found one of two local minima for the pressure drop across the nozzle. \rev{Because the optimizer
is deterministic, this dependence on the initial value indicates non-uniqueness
of the optimum and the presence of multiple local minima rather than stochastic scatter.} Each optimization run was done for 15 iterations with an initial search radius of $10\degree$ for the COBYQA algorithm. One solution has a lower optimal angle between $50.5\degree$ and $56.3\degree$, depending on the material, and the other solution has a larger optimal angle between $75.1\degree$ and $90\degree$. The flow fields with streamlines for both solutions for PLA are shown in Figure~\ref{fig:PLA-flow-field-angle} and for PET-G are shown in Figure~\ref{fig:PET-G-flow-field-angle}. Both materials behave very similar, which can be explained by the fact that the values of the relaxation time and the mobility factor are similar for both materials. The lower optimal angle for PLA and PET-G is $51.7\degree$ (Figure~\ref{fig:PLA-bild1-flow-field}) and $50.5\degree$ (Figure~\ref{fig:PET-G-bild1-flow-field}), respectively. The larger optimal angle is $75.8 \degree$ (Figure~\ref{fig:PLA-bild2-flow-field}) and $75.1 \degree$ (Figure~\ref{fig:PET-G-bild2-flow-field}) for both materials. 
The third material PA6/66 (Figure~\ref{fig:PA6-66-flow-field-angle}), behaves a bit differently, as it has a higher relaxation time and a higher mobility factor than the other two materials. Here, the lower optimal angle is $56.3\degree$ (Figure~\ref{fig:PA6-66-bild1-flow-field}), and the larger optimal angle is at $90\degree$ (Figure~\ref{fig:PA6-66-bild2-flow-field}). Notice that here the maximum extrusion velocity is lower compared to the other two materials, despite the same average extrusion velocity. This is due to the higher shear thinning effects of PA6/66 due to the larger mobility factor $\alpha_G$.\\
For all materials, the lower-angle optimum is characterized by vortex-free flow, whereas at the higher-angle optimum, a pronounced recirculation region forms in the upper corner of the nozzle, as indicated by the streamline patterns. Notably, the configuration with recirculation exhibits a lower overall pressure drop.
The lower pressure drop for the larger angle can be explained by the fact that the friction losses between the vortex and the main flow are lower than the direct contact of the flow with the wall. In practice, however, these recirculation areas should be avoided because they can promote backflow in the nozzle or lead to material degradation in the vortex. \rev{The material degradation happens, because the polymer is trapped in high-shear, high-temperature regions for extended times, causing thermal and mechanical chain scission.} Therefore, despite the slightly lower pressure drop of the case with the larger angle, the solution without the vortex is preferable. For angles in between the two optimal values, a smaller vortex is present, which increases the pressure drop, as the vortex does not cover the whole upper corner of the nozzle.
This issue could be addressed in future work by modifying the objective function to explicitly penalize vortices. In this work, the presence of vortices is checked manually by observing the streamlines.
It can also be observed that for all materials, the streamlines for the case without the vortex look similar to the case with the larger angle, when the vortex area would have been cut off.\\
As a summary, the results from the angle optimization for an extrusion velocity of 110 mm/s are shown in Table~\ref{tab_angle_study_results}. The relative improvements are also shown; they are around 4–5\% for all materials when the optimal angle is compared to the base case with a contraction angle of $30\degree$. The relative improvement for the case with the vortex is above 5\% for all materials, which is higher than the relative improvement for the case without vortices. However, because vortices are generally undesirable, in the following results for angle and spline-based shape optimization, we only consider solutions where no vortices are present.\\
\begin{table}[!htbp]
\centering
\caption{Angle optimization results for the extrusion velocity of 110 mm/s.}
\label{tab_angle_study_results}

\begin{tabular}{l l S S S }
\toprule
Material & Optimization case & {Angle[$\degree$]} & {$\Delta p$}[MPa] & {relative improvement[\%]} \\
\midrule
PLA   & base case &  30  & 2.0120 & \text{-} \\
PLA   & without vortex & 51.7  & 1.9218 & 4.48 \\
PLA   & with vortex & 75.8  & 1.9058 & 5.28 \\
\midrule
PET-G   & base case &  30 & 0.5282 & \text{-} \\
PET-G   & without vortex & 50.5 & 0.5048 & 4.43 \\
PET-G   & with vortex & 75.1  & 0.5011 & 5.13  \\
\midrule
PA6/66   & base case &  30  & 1.1798 & \text{-}\\
PA6/66   & without vortex & 56.3  & 1.1237 & 4.76 \\
PA6/66   & with vortex &  90 &  1.1153 & 5.47 \\

\botrule
\end{tabular}
\end{table}
Figure~\ref{fig:angle-convergence} shows the convergence behavior of the optimizer for the vortex-free solutions. For comparison across materials, the pressure drop is normalized by its initial value corresponding to a contraction angle of $30^\degree$. For all materials considered, the normalized pressure drop decreases smoothly over the first five optimization iterations, followed by only minor adjustments in subsequent iterations. Convergence is achieved within ten iterations for all materials, beyond which no further significant reductions in pressure drop are observed.\\
\begin{figure}[!htbp]
    \centering
    \begin{minipage}{0.5\textwidth}
        \centering


    \begin{tikzpicture}
        \begin{axis}[
            width=\linewidth,
            grid=major,
            xlabel={Number of iterations[-]},
            ylabel={Relative pressure drop [\%]},
            yticklabel style={
                /pgf/number format/fixed,
                /pgf/number format/precision=5
            },
            legend style={
                at={(0.5,1.05)},
                anchor=south,
                legend columns=3
            },
            ticklabel style={font=\small},
            label style={font=\small},
            legend style={font=\small},
            line width=1pt,
            tick style={thick},
        ]

    \pgfmathsetmacro{\initval}{2012.0282584564434*0.01}
        \addplot[color=red, mark=square*] coordinates {
        (0,{2012.0282584564434/\initval})
        (1,{1968.41198667158/\initval})
        (2,{1942.0684667406174/\initval})
        (3,{1927.744684135502/\initval})
        (4,{1923.916487531676/\initval})
        (5,{1923.8368667055602/\initval})
        (6,{1923.0025626022345/\initval})
        (7,{1921.826311187972/\initval})
        (8,{1924.770762605679/\initval})
        (9,{1921.8406613159632/\initval})
        };
        \addlegendentry{PLA}
        
\pgfmathsetmacro{\initval}{528.1784725792463*0.01}
        \addplot[color=black, mark=x] coordinates {
(0,{528.1784725792463/\initval})
(1,{517.1294028978253/\initval})
(2,{510.35406505854615/\initval})
(3,{506.60913243986175/\initval})
(4,{505.5336091365273/\initval})
(5,{505.07590912350133/\initval})
(6,{504.81219576000044/\initval})
(7,{504.54848239649954/\initval})
(8,{504.60941148145396/\initval})
(9,{504.6703405664084/\initval})
(10,{504.7312696513628/\initval})
(11,{504.7921987363172/\initval})
        };
        \addlegendentry{PET-G}

\pgfmathsetmacro{\initval}{1179.8203912744514*0.01}

        \addplot[
            color=blue, mark=o
        ] coordinates {
            (0,{1179.8203912744514/\initval})
            (1,{1156.2412577213322/\initval})
            (2,{1141.461983757516/\initval})
            (3,{1132.3504752257404/\initval})
            (4,{1126.5511483770533/\initval})
            (5,{1124.2913172384715/\initval})
            (6,{1123.7339008299339/\initval})
            (7,{1123.8428143533426/\initval})
            (8,{1123.7192975660232/\initval})
            (9,{1123.8504239567283/\initval})
            (10,{1124.3735932574386/\initval})
        };
        \addlegendentry{PA6/66}
        \end{axis}
    \end{tikzpicture}


    \end{minipage}
    \caption{Convergence of the relative pressure drop for the angle optimization without vortices for each material with an extrusion velocity of 110 mm/s. Source: Authors own work}
    \label{fig:angle-convergence}
\end{figure}
Next, angle-based optimization was performed for extrusion velocities ranging from 50 mm/s to 140 mm/s. Only solutions corresponding to the lower-angle local minimum without recirculation were considered. The results are summarized in Figure~\ref{fig:viscoelastic-angle-optimization}. The left panel shows the optimal nozzle angle as a function of extrusion velocity (Figure~\ref{fig:viscoelastic-angle-optimization_angle}), while the right panel reports the corresponding relative pressure-loss reduction (Figure~\ref{fig:viscoelastic-angle-optimization_improvement}).
For PLA and PET-G, the optimal angle remains within a narrow range of approximately $48\degree$–$52\degree$. The relative improvement is nearly constant for PLA at about 4.3\%, whereas for PET-G it decreases with increasing extrusion velocity, from 5.2\% to 4.4\%. In contrast, PA6/66 exhibits a stronger sensitivity to extrusion velocity, with the optimal angle increasing from $54.4\degree$ at 50 mm/s to $63.5\degree$ at 140 mm/s, accompanied by an increasing relative improvement. These results indicate that the extent to which nozzle geometry must be adapted to printing speed depends strongly on the material rheology.
\begin{figure}[!htbp]
\begin{subfigure}{0.5\textwidth}
    \centering


    \begin{tikzpicture}
        \begin{axis}[
            width=\linewidth,
            grid=major,
            xlabel={Extrusion velocity [mm/s]},
            ylabel={Optimal angle $\alpha$ [$\degree$]},
            legend style={
                at={(0.5,1.05)},
                anchor=south,
                legend columns=3
            },
            ticklabel style={font=\small},
            label style={font=\small},
            legend style={font=\small},
            line width=1pt,
            tick style={thick},
            every axis plot/.append style={thick, mark=*}
        ]
        \addplot[color=red, mark=square*] coordinates {
            (50,48.86249978)
            (80,51.67585236)
            (110,51.67787871)
            (140,48.84125208)
        };
        \addlegendentry{PLA}

        \addplot[color=black, mark=x] coordinates {
            (50,51.12866725)
            (80,48.73068828) 
            (110,50.49508107) 
            (140,48.76552734)
        };
        \addlegendentry{PET-G}

        \addplot[color=blue, mark=o] coordinates {
            (50,54.41464469)
            (80,57.18789124)
            (110,56.3310009)
            (140,63.46864543)
        };
        \addlegendentry{PA6/66}

        \end{axis}
    \end{tikzpicture}

%
    \caption{Optimal angle}
    \label{fig:viscoelastic-angle-optimization_angle}
\end{subfigure}
\begin{subfigure}{0.5\textwidth}
    \centering

    \begin{tikzpicture}
        \begin{axis}[
            width=\linewidth,
            grid=major,
            xlabel={Extrusion velocity [mm/s]},
            ylabel={Relative pressure drop improvement [\%]},
            legend pos=north east,
            ticklabel style={font=\small},
            yticklabel style={/pgf/number format/fixed, /pgf/number format/precision=4},
            label style={font=\small},
            legend style={font=\small, yshift=40pt},
            line width=1pt,
            tick style={thick},
            every axis plot/.append style={thick, mark=*}
        ]

        \addplot[color=blue, mark=o] coordinates {
            (50,100-538.6065775971899/561.0126960838616 * 100)
            (80,100-831.9894240025994/871.0665750360197 * 100)
            (110,100-1123.7192975660232/1179.8203912744514 * 100)
            (140,100-1411.8603201797764/1487.7211471167025 * 100)

        };
        
        \addplot[color=black, mark=x] coordinates {
            (50,100-245.50750158855965/256.82726881318956 * 100)
            (80,100-376.3754043943601/393.54241519159825 * 100)
            (110,100-504.7921987363172/528.1784725792463 * 100)
            (140,100-625.6142344391297/653.8940511965941 * 100)
        };

        \addplot[color=red, mark=square*] coordinates {
            (50,100-938.8571703916261/990.3160690768017 * 100)
            (80,100-1437.794930391227/1508.3223287870599 * 100)
            (110,100-1921.826311187972/2012.0282584564434 * 100)
            (140,100-2397.125550072087/2508.293858902563 * 100)
        };

        \end{axis}
    \end{tikzpicture}

    \caption{Relative improvement of the pressure drop}
    \label{fig:viscoelastic-angle-optimization_improvement}
\end{subfigure}
\caption{Optimal angle (a) and relative pressure drop improvement (b) for different extrusion velocities for PLA, PET-G, and PA6/66 material. Source: Authors own work}
\label{fig:viscoelastic-angle-optimization}
\end{figure}

\FloatBarrier

\subsection{Spline Parametrization}\label{Spline Parametrization}
\begin{figure}[!htbp]
    \centering
    {\centering
      \includegraphics[width=0.5\textwidth]{plots_and_figures/flow-field-viscoelastic/PLA-legend.pdf} 
      \\
      }
    \begin{subfigure}[b]{0.9\textwidth}
        \includegraphics[width=\linewidth]{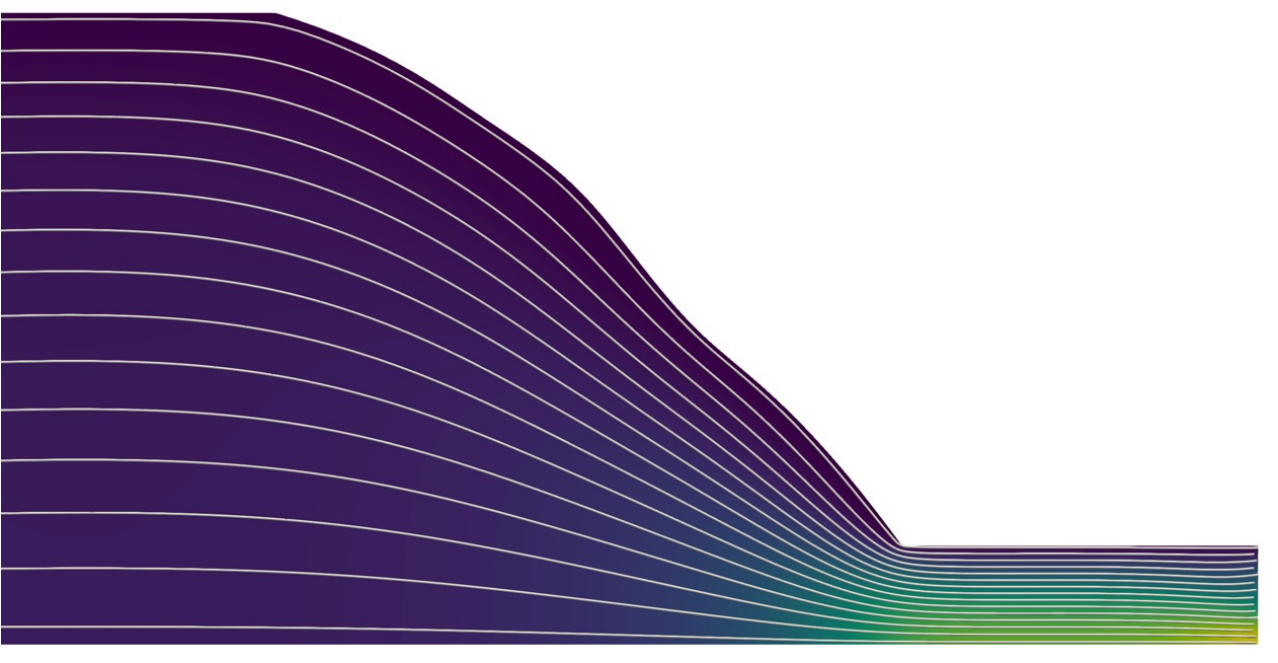}
    \end{subfigure}
   \caption{Flow field of the velocity in x-direction (u \rev{[$\si{\milli\meter\per\second}$]}) with the streamlines for the spline-based optimal shape for PLA material. The extrusion velocity is $110$~mm/s. Source: Authors own work} 
   \label{fig:PLA-flow-field-spline}
\end{figure}

\begin{figure}[!htbp]
    \centering
    {\centering
      \includegraphics[width=0.5\textwidth]{plots_and_figures/flow-field-viscoelastic/PET-G-legend.pdf} 
      \\
      }
    \begin{subfigure}[b]{0.9\textwidth}
        \includegraphics[width=\linewidth]{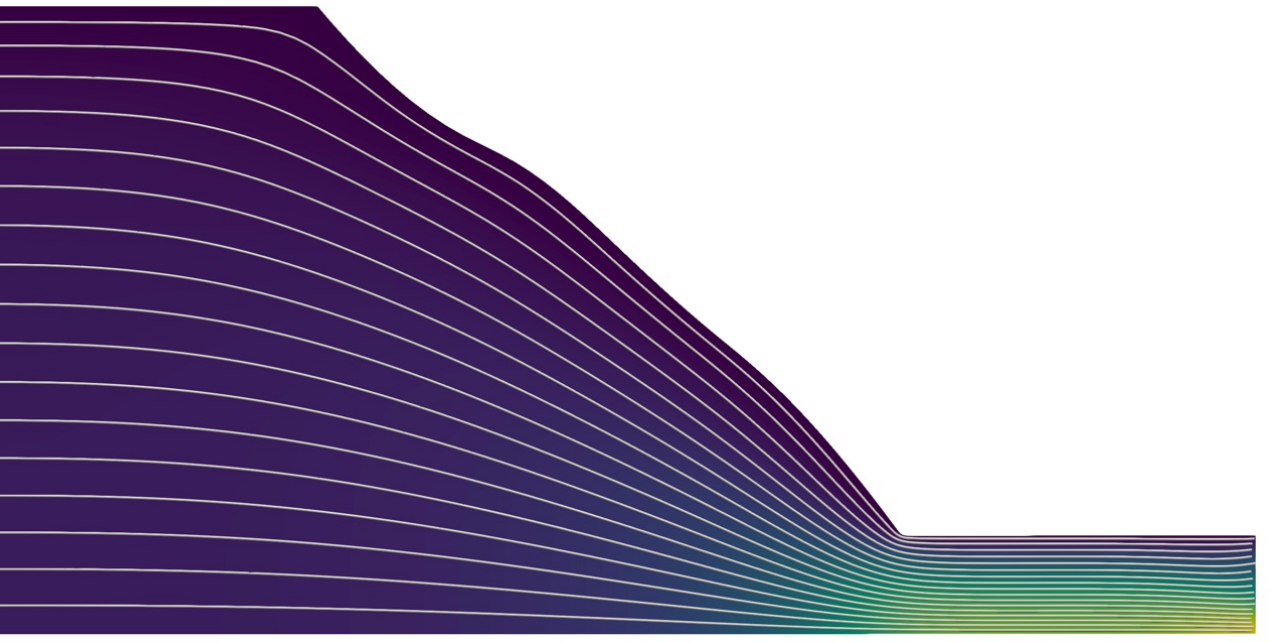}
    \end{subfigure}
   \caption{Flow field of the velocity in x-direction (u \rev{[$\si{\milli\meter\per\second}$]}) with the streamlines for the spline-based optimal shape for PET-G material. The extrusion velocity is $110$~mm/s. Source: Authors own work} 
   \label{fig:PET-G-flow-field-spline}
\end{figure}

\begin{figure}[!htbp]
    \centering
    {\centering
      \includegraphics[width=0.5\textwidth]{plots_and_figures/flow-field-viscoelastic/PA6-66-legend.pdf} 
      \\
      }
    \begin{subfigure}[b]{0.9\textwidth}
        \includegraphics[width=\linewidth]{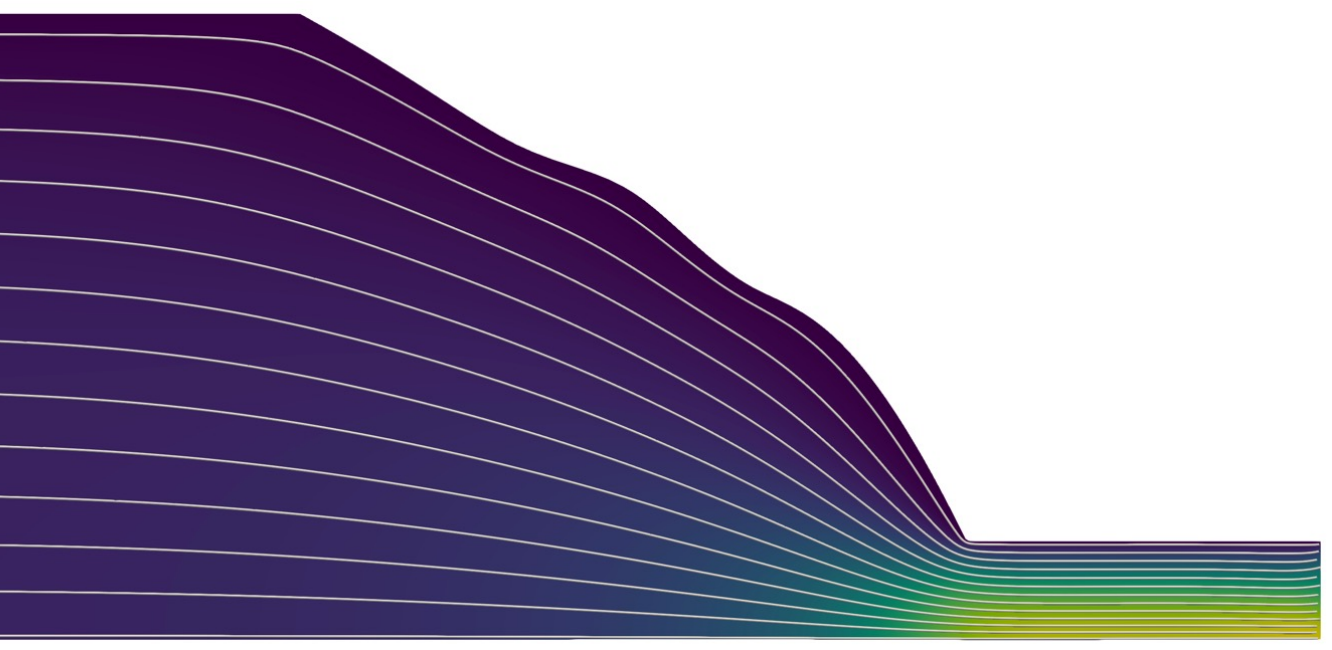}
    \end{subfigure}
   \caption{Flow field of the velocity in x-direction (u \rev{[$\si{\milli\meter\per\second}$]}) with the streamlines for the spline-based optimal shape for PA6/66 material. The extrusion velocity is $110$~mm/s. Source: Authors own work} 
   \label{fig:PA6/66-flow-field-spline}
\end{figure}

This section presents the results obtained using the spline-based geometric parametrization. Unless stated otherwise, all simulations were performed at an extrusion velocity of 110 mm/s. The optimizer ran for each case for around 100-160 iterations until convergence. In preliminary numerical experiments, the optimizer settings were tuned, in particular the initial search radius and the relative scaling between the spline control point coordinates and the nozzle angle. The selected initial search radii for the COBYQA algorithm were 0.8 mm for the control point positions and $6^\circ$ for the angle.
Subsequently, a study was conducted to assess the influence of the number of spline control points on the optimization outcome. Shape optimization was performed using 3, 5, 7, and 9 control points for the PA6/66 material. The results, shown in Figure~\ref{fig:spline-point-numbers}, indicate that the lowest pressure drop is obtained with five control points. Although the differences are relatively small, five control points were used in all subsequent spline-based shape optimization studies.
\begin{figure}[!htbp]
    \centering
    \begin{minipage}{0.5\textwidth}
        \centering


    \begin{tikzpicture}
        \begin{axis}[
            width=\linewidth,
            grid=major,
            xlabel={Number of spline control points[-]},
            ylabel={Pressure drop [MPa]},
            yticklabel style={
                /pgf/number format/fixed,
                /pgf/number format/precision=5
            },
            legend style={
                at={(0.5,1.05)},
                anchor=south,
                legend columns=3
            },
            xtick={3,5,7,9},
            ticklabel style={font=\small},
            label style={font=\small},
            legend style={font=\small},
            every axis plot/.append style={thick, mark=*}
        ]

        \addplot[color=blue, mark=o,only marks] coordinates {
            (3,1.1195)
            (5,1.1166)
            (7,1.1182)
            (9,1.1186)
        };

        \end{axis}
    \end{tikzpicture}


    \end{minipage}
    \caption{Pressure drop for the shape optimization with spline-based parametrization with different numbers of control points for PA6/66 with an extrusion velocity of 110 mm/s. Source: Authors own work}
    \label{fig:spline-point-numbers}
\end{figure}
Figures~\ref{fig:PLA-flow-field-spline} and~\ref{fig:PET-G-flow-field-spline} show the optimized nozzle geometries obtained using the spline-based parametrization for PLA and PET-G, respectively. Compared to the angle-optimized configurations without recirculation, the overall nozzle angle is slightly reduced for both materials. In the vicinity of the re-entry corner, the local angle remains comparable to that obtained from angle-based optimization.
Further upstream, the spline-based shapes deviate from the purely angular geometry. For PLA, the profile becomes flatter and more smoothly rounded, whereas for PET-G the geometry first becomes flatter but then bends more sharply upward toward the upper part of the nozzle. The corresponding streamline plots indicate smooth flow without the formation of recirculation regions for either material.
For PA6/66, the spline-optimized geometry differs more noticeably from the corresponding angle-optimized shape, as shown in Figure~\ref{fig:PA6/66-flow-field-spline}. The geometry exhibits a relatively steep angle near the re-entry corner, followed by a more pronounced flattening upstream compared to PLA and PET-G. In addition, two small local shape variations are present along the spline profile. The streamline patterns indicate smooth flow without the formation of recirculation regions.
Table~\ref{tab_spline_study_results} compares the results of angle-based and spline-based shape optimization. For all three materials, the spline-based parametrization yields a modest additional reduction in pressure drop compared to the angle-optimized configuration without recirculation. For PLA, the pressure drop decreases from 1.9218 MPa in the angle-optimized case to 1.9154~MPa with spline-based optimization, corresponding to an increase in relative improvement from 4.48\% to 4.80\%. A similar trend is observed for PET-G, where the pressure drop is reduced from 0.5048 MPa to 0.5024 MPa, increasing the relative improvement from 4.43\% to 4.88\%. The largest additional improvement is obtained for PA6/66, for which the pressure drop decreases from 1.1237~MPa to 1.1166~MPa and the relative improvement increases from 4.76\% to 5.36\%. 
Figure~\ref{fig:spline-convergence} illustrates the convergence behavior of the spline-based optimization for the three materials considered. The relative pressure drop is shown as a function of the optimization iteration. During the first 20–30 iterations, depending on the material, the pressure drop decreases rapidly as the COBYQA algorithm explores the design space. This initial phase is followed by a regime of gradually diminishing improvements up to approximately 80 iterations. Beyond roughly 100 iterations, the optimization reaches convergence for all materials, with no further significant reductions in pressure drop observed.
Compared to the angle-based optimization, the spline-based approach requires approximately an order of magnitude more iterations to converge, reflecting the increased dimensionality of the design space.\\
\begin{table}[!htbp]
\centering
\caption{Comparison of the angle and spline-based shape optimization for the extrusion velocity of 110 mm/s.}
\label{tab_spline_study_results}
\begin{tabular}{l l S S S }
\toprule
Material & Optimization case & {Angle[$\degree$]} & {$\Delta p$}[MPa] & {relative improvement[\%]} \\
\midrule

PLA   & angle & 51.7  & 1.9218 & 4.48 \\
PLA   & spline & \text{-}  & 1.9154 & 4.80 \\
\midrule
PET-G   & angle & 50.5 & 0.5048 & 4.43 \\
PET-G   & spline & \text{-}  & 0.5024 &  4.88 \\
\midrule
PA6/66   & angle & 56.3  & 1.1237 & 4.76 \\
PA6/66   & spline &  \text{-} & 1.1166& 5.36 \\

\botrule
\end{tabular}
\end{table}
\begin{figure}[!htbp]
    \centering
    \begin{minipage}{0.7\textwidth}
        \centering
        \include{plots_and_figures/spline-convergence}

    \end{minipage}
    \caption{Convergence of the relative pressure drop for the spline-based shape optimization for each material with an extrusion velocity of 110 mm/s. Source: Authors own work}
    \label{fig:spline-convergence}
\end{figure}
Overall, these results indicate that spline-based parametrization marginally exceed the reductions in pressure loss obtained with the angle parameterization while increasing the geometric complexity of the nozzle. Further investigation is required to assess whether such increased shape complexity is justified in terms of manufacturability and associated costs, given the relatively marginal gains in pressure drop.

\FloatBarrier
\section{Discussion and Outlook}\label{sec12}
This work investigated the influence of nozzle geometry on pressure loss in fused deposition modeling, with a focus on improving high-speed printing performance through shape optimization. A flexible optimization framework was developed that supports both angle-based and spline-based geometric parametrizations and incorporates a viscoelastic Giesekus model to capture polymer melt behavior.
The results demonstrate that nozzle geometry has a measurable impact on pressure loss and that angle-based optimization already achieves most of the attainable improvement. For the materials considered, angle optimization revealed the existence of multiple local optima, including configurations with recirculation regions that yield lower pressure loss but are undesirable due to increased residence times and the associated risk of material degradation and nozzle clogging. 
Spline-based parametrization provides additional geometric flexibility and yields modest further reductions in pressure loss, particularly for materials exhibiting stronger sensitivity to nozzle shape, such as PA6/66. However, these improvements remain minor when compared to the increased geometric complexity introduced by spline-based designs. This highlights an important trade-off between performance gains and manufacturability. \rev{The present results suggest that within the investigated parameter range of the planar model, a vortex-free contraction angle of roughly $50\degree$--$56\degree$ already captures most of the achievable pressure-loss reduction, whereas spline-based refinements provide only marginal additional benefit. The conclusions should therefore be interpreted as design guidance for comparative shape optimization within the simplified model class considered here.} Overall, the findings suggest that simple, angle-optimized nozzle geometries represent a robust and manufacture-ready solution for the materials and conditions used in this study, while more complex spline-based designs may be justified only in specific cases. \\
\rev{Future work will therefore focus on broadening the physical realism of the model, in particular by considering axisymmetric geometries, thermal effects, and polymer melting, and by comparing the optimized shapes against dedicated experimental measurements of extrusion force and temperature.}
\backmatter

\bmhead{Acknowledgments}
The presented investigations were carried out at RWTH Aachen University within the framework of the Collaborative Research Centre SFB1120-236616214 “Bauteilpräzision durch Beherrschung von Schmelze und Erstarrung in Produktionsprozessen” and funded by the Deutsche Forschungsgemeinschaft (DFG, German Research Foundation). The work was also funded by DFG project 566317635. The sponsorship and support are gratefully acknowledged.
The authors also gratefully acknowledge the German Federal Ministry of Research, Technology and Space (BMFTR), and the state of North Rhine-Westphalia for supporting this work as part of the NHR funding.
Additionally, the authors gratefully acknowledge the computing time provided to them by the NHR Center NHR4CES at RWTH Aachen University on the CLAIX high-performance computing cluster (project number p0024828).
This work was also partly supported by the IRTG Modern Inverse Problems which is funded by the DFG – 333849990/GRK2379.

\bmhead{Data Availability}
The data and materials for this publication are available on request at the following link \url{http://hdl.handle.net/21.11102/0adeb604-8a42-46b1-a734-85d6bc56fc16}.

\FloatBarrier



\bibliography{sn-bibliography}


\end{document}